\documentclass[prb,reprint,nofootinbib,superscriptaddress]{revtex4-2}
\usepackage{amsmath,amssymb}
\usepackage{graphicx,bm}
\usepackage[pdfencoding=auto,colorlinks=true,linkcolor=blue]{hyperref}
\usepackage{lmodern}
\usepackage{mathtools}
\usepackage{empheq}
\usepackage{float}
\usepackage{color}

\newcommand{\sgn}{\mathrm{sgn}}
\newcommand{\bea}{\begin{eqnarray}}
\newcommand{\eea}{\end{eqnarray}}
\newcommand{\beq}{\begin{equation}}
\newcommand{\eeq}{\end{equation}}
\newcommand{\be}{\begin{equation}}
\newcommand{\ee}{\end{equation}}
\newcommand{\ba}{\begin{align*}}
\newcommand{\ea}{\end{align*}}

\renewcommand{\(}{\left(}
\renewcommand{\)}{\right)}
\renewcommand{\[}{\left[}
\renewcommand{\]}{\right]}

\newcommand{\non}{\nonumber}
\newcommand{\kpara}{k_{\|}}

\begin{document}
\title{BCS-like $T_c$ does not necessarily imply BCS pairing mechanism: the case of magnetically-mediated quantum-critical pairing}
 \author{Yuxuan Wang }
\affiliation{Department of Physics, University of Florida, Gainesville, Florida 32601}

\author{Andrey V. Chubukov}
 \affiliation{School of Physics and Astronomy and William I. Fine Theoretical Physics Institute, University of Minnesota, Minneapolis, MN 55455, USA}

\begin{abstract}
In the BCS theory of superconductivity, an instability towards pairing develops at arbitrary weak dimensionless coupling
  $\lambda$ due to a divergence of logarithmic  perturbative series for the pairing susceptibility (Cooper logarithms) at  $T_c \sim \omega_0 e^{-1/\lambda}$, where $\omega_0$ is an energy cutoff.
   On the contrary,  in many models of superconductivity out of a non-Fermi liquid, Cooper logarithm is absent and superconductivity emerges only when $\lambda$ exceeds a certain threshold. We argue that there are situations when there is no threshold and at weak coupling the formula for $T_c$ is BCS-like, yet the
    origin of the
    pairing instability is  fundamentally different from that in the BCS scenario.
    As an example, we revisit superconductivity in a metal at the onset of $(\pi,\pi)$  spin-density-wave order.  Earlier studies of this problem found no threshold and a BCS-like expression for $T_c$ at weak coupling.  We argue that, despite this, the pairing is not caused by the Cooper logarithm  and in many respects is  qualitatively similar to that in non-Fermi liquids.
\end{abstract}
\maketitle

\section{Introduction}

Non-BCS pairing of fermions  from a non-Fermi liquid normal state has attracted substantial interest in recent years in both condensed-matter and high-energy communities.
 In a nutshell, pairing out of a non-Fermi liquid is
 mediated by a massless boson, the same one that destroys Fermi liquid behavior. This leads to  non-trivial
  competition between two opposite tendencies,  one towards pairing and the other towards a non-Fermi liquid. In some cases, this competition leads to
   superconductivity with enhanced critical temperature $T_c$~\cite{color1,acf1,WC,Mross2015,raghu-enhanced,first-mats-2,efetov}, while in others
    it leads to a complete suppression of pairing
     such that
     the system remains metallic down to zero temperature~\cite{acf1,raghu2015metallic,ysyk}.         In both cases,         strong pairing fluctuations  above $T_c$        give rise  to a pseudogap behavior~\cite{first-mats-2,first-mats-3,gap-filling-QMC}.

In a Fermi liquid with a weak attractive interaction,  pairing instability develops via the well-known BCS mechanism. Namely,  the pairing susceptibility $\chi_{\rm pp}$ is expressed within perturbation theory via a power series of a Cooper logarithm --- the term $\lambda\log( \omega_0/T)$, where $\lambda$ is a dimensionless coupling, $\omega_0$ is an energy cutoff and $T$ is the temperature. The series is geometrical and yields  $\chi_{\rm pp}  \propto 1/[1-\lambda \log(\omega_0/T)]$, which diverges at $T=T_c \sim \omega_0 e^{-1/\lambda}$,
 signaling the onset of superconductivity.

By contrast, in non-Fermi liquids, such as quantum-critical metals~\cite{acf1,raghu2015metallic,first-mats-1,first-mats-2,first-mats-3,wang-wang-torroba,WC,Yuzbashyan1}
 and  Yukawa-SYK models~\cite{ysyk,ysyk_esterlis,ysyk2,ysyk_pan, Wang_2021,ysyk_hauck, ysyk_classen, ysyk_valentinis, ysyk_choi, ysyk_inkof, ysyk_li} of dispersion-less fermions with random interactions,
 perturbation series for $\chi_{\rm pp}$
  are rather peculiar: they are
     logarithmic, as in
      BCS theory,
       but
     the
     argument of the logarithm  depends on the  running frequency of a fermion, $\omega$, and the combinatorial factor
       for the $n$-th  term in the series is $1/n!$ rather than $1$. The
        series  $ \lambda^n |\log{(\omega_0/|\omega|)}|^n/n!$ sum up into
       $\chi_{\rm pp}  (\omega) \propto \left(\omega_0/|\omega|\right)^\lambda$, which does not
        diverge
        at any finite frequency, i.e., any finite $T$.\footnote{Within the renormalization group (RG) framework
        the coupling constant $g$ in the Cooper channel flows
        to a  fixed point instead of running to infinity.}
 To search for a potential pairing instability, one has to go beyond the expansion in powers of the logarithms and compute the pairing susceptibility exactly. The outcome of the exact analysis is that the pairing instability develops, but only if $\lambda$ exceeds a certain threshold $\lambda_c$,  when nonlogarithmic corrections change the exponent for $\chi_{\rm pp}  (\omega)$
     to the extent that it becomes complex and   $\chi_{\rm pp}(\omega)$ develops oscillations.
Such behavior cannot be captured within
order-by-order expansion and
indicates a breakdown of perturbation theory in the particle-particle channel.  Solving the non-linear gap equation for the gap function $\Delta (\omega)$, one then finds that this breakdown implies an instability towards superconductivity.  This
 is often referred to as the {\it complex exponent} scenario of the pairing rather than the one based on
the summation of
   the Cooper logarithms.\footnote{In the RG language,  the  fixed points for $g$ formally
   move to the complex plane
   and become inaccessible~\cite{son_1,*son_b,*son_3,raghu2015metallic,ysyk}, and instead $g$ flows all the way to infinity. A similar scenario has been proposed to describe
the pseudo-critical behavior at putative deconfined quantum phase transitions~\cite{walking1,walking2}.}

There is one additional aspect in which this pairing differs from BCS. Namely, the non-linear gap equation has an infinite number of topologically distinct solutions, which differ in the number of oscillations of in  $\Delta_n(\omega)$.
  Each solution has its own $T_{c,n}$. The largest $T_{c,0}$ is for non-oscillating $\Delta_n(\omega)$, which is a global minimum of the condensation energy. Other solutions are saddle points.  In BCS theory, there exists a single solution for a superconducting order parameter at $T < T_c$ (the same holds in the Eliashberg theory for pairing out of a Fermi liquid~\cite{paper_2}).

 In a recent paper~\cite{Ojajarvi2024},
  Ojajarvi et al analyzed unconventional superconductivity in a 2D system with Hubbard-like repulsion $U$, near a single Van Hove point. They found that $p$-wave superconductivity develops already for an arbitrary small $U$, and that $T_c \propto e^{-1/\lambda}$, where $\lambda \propto U$.  On the surface, this looks like BCS pairing.  The authors of ~\cite{Ojajarvi2024} however argued that there is no Cooper logarithm for this problem, and the exponential dependence of $T_c$ on $\lambda$ is due to logarithmical singularity in the density of states at the Van Hove point.  They argued that the pairing mechanism is very similar to the one for non-BCS pairing mediated by a massless boson. In particular, the pairing instability is the consequence of oscillations in $g(E)$, and there exists an infinite number of $T_{c,n}$ with topologically distinct gap function, of which $T_{c,0}$ is the largest.

  In this paper, we argue that the same holds for $d$-wave pairing in a metal near the onset of a spin-density-wave order with momentum $(\pi,\pi)$. 
   This pairing
   has been extensively analyzed in the context of
  cuprate and pnictide superconductors and other materials.  At and above optimal doping, the cuprates display  metallic behavior and the Fermi surface area matches Luttinger count, and magnetic-mediated pairing is often thought to be confined to the area near hot spots --- $\bm k_F$ points for which $\bm k_F + (\pi,\pi)$  is also on the Fermi surface.  The fermionic self-energy $\Sigma (\omega)$ right at a hot spot is singular, scaling as $\omega^{1/2}$. Such pairing of hot fermions then falls into ``pairing out of  non-Fermi liquid'' category, in which superconductivity only develops when the pairing interaction is above the threshold.
   However, for lukewarm fermions away from hot spots, the self-energy $\Sigma (\omega_m, k_{\parallel}) \propto \omega/|k_{\parallel}|$~\cite{ms1},
    where $k_{\parallel}$ is momentum deviation from a hot spot along the Fermi surface.
   Intuitively, as these fermions retain Fermi liquid behavior at any  $k_\|$, they can pair at arbitrary weak $d$-wave interaction $\lambda$ and, by proximity, impose pairing of  hot fermions.  Indeed, explicit calculations show~\cite{WC}
   that $T_c \propto e^{-1/\lambda}$, i.e., that it has the same functional form as in BCS theory.
   Although the result looks like BCS pairing of lukewarm fermions, we argue that this is not the case, and, despite
    the similarity with the BCS formula for $T_c$, the pairing mechanism is fundamentally non-BCS and is rather similar to the previous case for fermions near a single van Hove point.

   To prove our point, we 
   note that although a lukewarm fermion has Fermi-liquid-like, linear in $\omega$ frequency dependence, the prefactor critically depends on the distance from a hot spot, so some non-Fermi liquid physics can be expected.
     We analyze the linearized gap equation for the pairing vertex $\Phi$ and show
    that the pairing at weak coupling
     falls into the ``complex exponent'' scenario.
     In fact, a perturbative analysis of the pairing susceptibility yields a power series of
     $\log^2(\Lambda/T)$, but their sum does not diverge~\cite{WC}. As a consequence of the complex exponents (oscillation of $\Phi$), there is an infinite set of $T_{c,n}$.
     We first convert the original integral equation into a differential form, and begins with an analysis of a simplified toy model, which can be straightforwardly solved. Then we analyze the full differential equation, find its exact solution, and show that the pairing mechanism
    remains the same as in the toy model. Finally, we show that this differential equation can be directly mapped to the renormalization group (RG) equation for the coupling constant in the Cooper channel. The RG formulation of our pairing problem makes it clear that the pairing mechanism is fundamentally different from the BCS mechanism.

    The structure of the paper is the following. In Sec. \ref{sec:2}
    we
     provide some background information about the
     model and present
     the  gap equation for pairing between
     lukewarm fermions, which
      will be the point of departure for our analysis.  Here we also show that the BCS-type iteration procedure based on the summation of the leading powers of logarithms yields no  pairing instability.
      In Sec. \ref{sec:3} we convert  the integral gap equation into a differential one    
      and analyze it. We start in  Sec.~\ref{sec:3a} with the  simplified differential equation, using which we show that the pairing mechanism is that same as for the pairing out of a non-Fermi liquid,  but $T_c$ is non-zero for arbitrary weak $\lambda$ and has BCS-like form $T_c \propto e^{-c/\lambda}$, where $c =
      \pi/4$.   In Secs.~\ref{sec:3b} and \ref{sec:3c} we analyze the full differential equation
      and show that the mechanism of the pairing remains the same as in the toy model and $T$ has the same form.
       at the smallest $\epsilon$.
       In  Sec.~\ref{sec:3d} we discuss a more accurate mapping of the integral equation into a differential one, which extends beyond previous conversions and also borrows some results from~\cite{WC}.  We argue that this gives an accurate result $T_c \propto e^{-1/\lambda}$
       (i.e., $c=1$). In Sec.~\ref{sec:4} we reformulate our pairing problem in the RG framework, which makes its fundamental difference with BCS clear. We show that the RG equation is fully equivalent to the differential equation for the pairing vertex.~\cite{wang-raghu-torroba-17}
      We present our conclusions in Sec.~\ref{sec:5}. For completeness, in Appendix~\ref{app:1} we follow Ref.~\cite{wang-raghu-torroba-17} and analyze the interplay between the differential gap equation,
       extracted from the Eliashberg theory, and the RG equation for the flow of the running coupling for two intensively studied dynamical models: the one with a logarithmic (color superconductivity) and power-law (the $\gamma$-model) dynamical interaction. 
        In Appendix \ref{app:2} we compare this work with the 2013 paper by the two of us on the same subject, Ref.~\cite{WC}.

\section{Background: Model and integral gap equation}
      \label{sec:2}

We follow earlier works~\cite{Monthoux2007,Scalapino_2012,acf1,acs1,ms1,acn} and analyze the pairing near an antiferromagnetic QCP within
  the semi-phenomenological spin-fermion model. The model assumes that antiferromagnetic correlations develop already at
   high energies, of order bandwidth, and mediate interactions between low-energy fermions.
  The static part of the spin-fluctuation propagator is treated as a phenomenological input from high-energy physics, but the
   the dynamical Landau damping part is self-consistently obtained within the model as it
     comes entirely from low-energy fermions~\cite{acf1,acs1,ms1}.
    We assume, as in these earlier works, that the Eliashberg approximation (no corrections to spin-fermion vertex and self-consistent one-loop approximation for the self-energy) is valid in a wide range of frequencies, which extends below  the ones relevant for the pairing.  For a detailed analysis of the validity of the Eliashberg theory for a spin-fermion model see, e.g., Refs. \cite{acs1,ms1,sslee_2018,Zhang2024}.

The
    action of the
     spin-fermion
     model is  given by~\cite{cps,acs1,ms1}
 \begin{eqnarray}
{\mathcal{S}} &=&-\int_{k}G_{0}^{-1}\left( k\right) \psi _{k,\alpha }^{\dagger }\psi _{k,\alpha }+\frac{1}{2}\int_{q}\chi _{0}^{-1}\left( q\right) \ \bm{{S}}_{q}\cdot \bm{{S}}_{-q}  \nonumber \\ &&+\int_{k,q}\psi _{k+q,\alpha }^{\dagger }\bm\sigma
_{\alpha \beta }\psi _{k,\beta }\cdot \bm{{S}}_{-q}.\
  \label{startac}
\end{eqnarray}
  where $\int_k$ stands for the integral over
 $\bm{{k}}$ and the sum over
  Matsubara frequencies, $G_{0}\left( k\right) = G_0 (\omega_m, \bm{ k}) = 1/[i\omega_m - \bm{ v}_{F,\bm{ k}}\cdot  (\bm{ k}-\bm{ k}_F)]$
  is the bare  fermion propagator, and
   $\chi _{0}\left( q\right) = \lambda/\bm{ q}^2 $
  is the static propagator of collective bosons at quantum criticality and $\bm{ q}$ is measured  with respect to
  $\bm{ Q}$.

\subsection{Hot and lukewarm fermions}

\begin{figure}[t]
    \includegraphics[width=0.8\linewidth]{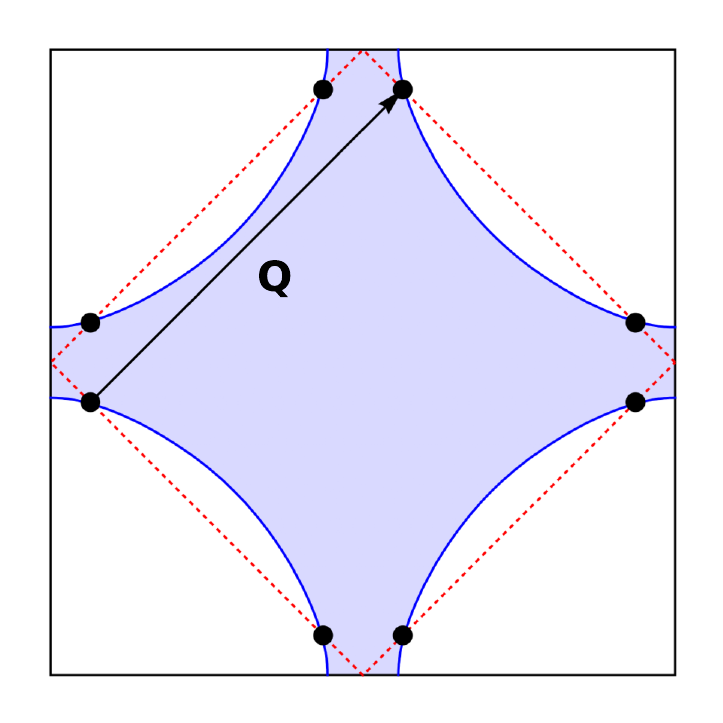}
    \caption{The Fermi surface near a quantum-critical point of antiferromagnetic order. The hot spots are labeled as black dots, connected by the wave vector $\bm Q$ of the order parameter.}
    \label{fig:FS}
\end{figure}

Antiferromagnetic fluctuations are peaked at
$\bm{ Q} = (\pi,\pi)$, and  mostly affect  fermions located near hot spots ---
     Fermi surface points $\bm{ k}_F$, for which $\bm{ k}_F + \bm{ Q}$ is also
      at the FS,
see Fig.~\ref{fig:FS}. The Fermi velocities at hot spots separated by $\bm{ Q}$ can be expressed as $\bm{ v}_{F,1} = (v_x, v_y)$
   and
   $\bm{ v}_{F,2} = (-v_x, v_y)$, where $x$ axis is along $\bm{ Q}$.

   Without loss of generality, we
   set
   $v_x = v_y$ in this work.

The fermion-boson coupling gives rise to fermionic and bosonic self-energies. In the normal state, bosonic self-energy accounts for Landau
damping of spin excitations, while fermionic self-energy accounts for  mass renormalization  and a finite lifetime of a fermion.  At one-loop level
  ~\cite{acs1,ms1}
  \bea
 &&\Sigma (\omega_m, k_\parallel) = \frac{3 \lambda}{4 \pi v_F}~\frac{2 \omega_m}{\sqrt{\gamma |\omega_m| +
 \kpara^2} + |\kpara|},   \label{2a}\\
 &&\chi (\Omega_m, \bm{ q}) = \frac{\lambda}{\bm{ q}^2 + |\Omega_m| \gamma}  \label{2}
 \eea
  where $\omega_m = \pi T (2m+1)$ is the fermionic Matsubara frequency, $\gamma =
  4 {\lambda}/(\pi v_F^2)$, and $k_\parallel$ is the deviation from a hot spot along the FS.
  The fermionic self-energy right at a hot spot has a non-FL form:
    \be
    \Sigma (\omega_m,0) =
   \sqrt{ \omega_0|\omega_m|} \sgn(\omega_m).
   \label{eq:hot}
   \ee
    where  $\omega_0 = 9 {\lambda}/16 \pi$.
   For fermions away from the hot spot, known as the ``lukewarm fermions'',
   $\Sigma (\omega_m, k_\parallel)$ retains a FL form at the smallest $\omega_m$ and scales as
    \be
    \Sigma (\omega_m, k_\parallel) \approx \sqrt{\gamma\omega_0}\,\frac{\omega_m}{2|k_\parallel|},~~~\kpara^2\gg \gamma\omega_m.
    \label{eq:warm}
    \ee

\subsection{Integral gap equation}
These normal state results are inputs for the analysis of the pairing.
The linear gap equation for pairing between fermions near two hot spots  has been obtained within Eliashberg approximation in \cite{WC}, and is
      \begin{align}
      \Phi(\kpara,\omega_m) =& \frac{\epsilon\lambda}{2v_F}  T\sum_{m'}\int\frac{d\kpara'}{2\pi} \frac{1}{|\omega'_m + \Sigma(\omega_m',\kpara')|}\non\\
      &\times\frac{ \Phi(\kpara',\omega_m')}{\kpara^2+\kpara'^{2}+\gamma|\omega_m-\omega_m'|}.
      \label{eq:gap}
      \end{align}

    The parameter $\epsilon$ is the relative strength of the spin-mediated  interactions in the $d$-wave particle-particle channel and $s-$wave particle-hole channel.   We follow Refs.~\cite{acf1,ms1,WC,first-mats-1} and treat $\epsilon$ as a small parameter\footnote{This can be justified by e.g., formally extending the model to a matrix large-$N$ theory~\cite{raghu2015metallic},
  in which case
  $\epsilon=1/N$~\cite{Chubukov_2020a,paper_1}. }
For simplicity, below we will refer to \eqref{eq:gap} as
the gap equation.

\subsubsection{Pairing out of non-Fermi liquid}

In the pioneering study of the pairing in the spin-fermion model~\cite{acf1}, Abanov et al. assumed that the pairing predominantly comes from hot fermions, for which the self-energy is given by
\eqref{eq:hot}.
Upon substituting this self-energy into \eqref{eq:gap}, setting $k_{\parallel} =0$ and integrating over $k^{'}_{\parallel}$,
  the gap equation takes the universal form\footnote{Eq. (\ref{ch_1}) is a specific realization of the
 pairing in the $\gamma$-model - a generic model of pairing out of a non-Fermi liquid by an effective $V(\Omega_m) \propto 1/|\Omega_m|^\gamma$;  the corresponding $\Sigma (\omega_m) \propto |\omega_m|^{1-\gamma}$.  Eq. (\ref{ch_1}) corresponds to $\gamma =1/2$.}
 \beq
 \Phi (\omega_m) = \frac{\epsilon T}{4} \sum_{\omega_{m'}} \frac{\sqrt{\omega_0}}{|\omega_{m'} +\Sigma (\omega_{m'})|} \frac{\Phi (\omega_{m'})}{|\omega_m- \omega'_m|^{1/2}}.
 \label{ch_1}
 \eeq

  The temperature, at which this equation has a non-zero solution,
   is the onset of
   the pairing.
   In the absence of strong phase fluctuations, this is superconducting $T_c$.

  The solution of the gap equation is the following~\cite{acf1,Chubukov_2020a,paper_1}.
 First,
  there is a threshold for the pairing, i.e., $T_c$ is finite only at $\epsilon > \epsilon_c= 0.22$.  Near the threshold, 
   $T_c \sim \omega_0 \exp\left[-b/(\epsilon-\epsilon_c)^{1/2}
   \right]$, $b \approx 3.41.$
(for similar results in other systems see \cite{son_1, son_b, Kaplan_2009}).
 Second,
  the pairing instability does not come from the summation of Cooper logarithms
   and is revealed only by going
   beyond the logarithmic approximation.
   Third, there is a tower of critical temperatures, $T_{c,n}$, which all emerge  once $\epsilon$ exceeds $\epsilon_c$ and scale as
   $\omega_0 \exp{\[-b_n/(\epsilon-\epsilon_c)^{1/2}
   \]}$.
   The $T_c$  given above is the largest $T_{c,0}$.  Each  $T_{c,n}$ signals an instability towards a state with  topologically distinct $\Phi (\omega_m)$.

\subsubsection{Pairing of lukewarm fermions}

 Subsequent studies~\cite{ms1,WC}, however, have found that the threshold at $\epsilon_c$ is actually a crossover rather than a sharp boundary, i.e., $T_c$ is finite even when $\epsilon$ is small. The argument is that lukewarm fermions in Eq.~\eqref{eq:warm} also contribute to pairing. Because these fermions display a Fermi liquid behavior, it seems natural to expect that the pairing of lukewarm fermions  falls into the   BCS framework.  Indeed, $T_c$, obtained in Ref.~\cite{WC}, remains finite even at the smallest $\epsilon$ and  has a familiar BCS form
     \beq
     T_c \sim \omega_0 e^{-1/\epsilon}.
\label{ch_2}
\eeq
We show below that the analogy with BCS is deceptive.  Namely, while there is no threshold, the pairing mechanism is still qualitatively different from BCS. In particular, we show  that there is a tower of $T_{c,n}$ of which the one in Eq. (\ref{ch_2}) is the largest.

The gap equation for lukewarm fermions is obtained from (\ref{eq:gap}) by using the self-energy from  (\ref{eq:warm}) and neglecting the dynamical Landau damping term in the bosonic propagator, which is
 smaller than the static term for frequencies and momenta, relevant to pairing, as one can verify a'posteriori.
      %
To see this, we focus on the lukewarm regime ($\gamma\omega_m'\ll\kpara'^2$) in the integral of Eq.~\eqref{eq:gap} and drop the $\omega_m'$ dependence in the bosonic propagator. From the remaining terms in the bosonic propagator, we obtain
that the pairing vertex in different regimes of $(\kpara,\omega_m)$  is described by a single-variable function:
\be
\Phi(\kpara, \omega_m) =\begin{cases}

\Phi(\kpara^2/\gamma)&\textrm{ if $\kpara^2\gg \gamma\omega_m$}\\
\Phi(\omega_m)&\textrm{ if $\kpara^2\ll \gamma\omega_m$.}\end{cases}
\ee
Plugging this
 into
 the right hand side of Eq.~\eqref{eq:gap}, we
  obtain
\begin{align}
\!\!\!
\Phi(\kpara^2/\gamma) =& \frac{\epsilon}{\pi} \int_{\gamma T}^{\gamma\omega_0}\frac{d\kpara^2}{\kpara^2+\kpara'^2} \log\({\kpara^2}/{\gamma T}\)
\Phi(\kpara'^2/\gamma) \label{eq:phi}\\
+&\frac{\epsilon\alpha\lambda}{\pi v_F} \int_{T}^{\omega_0}\frac{d\omega_m'}{\kpara^2+\gamma\omega'_m}\int_{0}^{\sqrt{\gamma\omega_m'}} \frac{d\kpara'}{\sqrt{\omega_0\omega_m'}}
\Phi(\omega_m'),\non
\end{align}
where $\alpha=\mathcal{O}(1)$.
 For small $\epsilon$, the contribution from the first line
 is the largest one.  Keeping only this term and
 introducing
 $x=\kpara^2/\gamma T$ and $y= \kpara'^2/\gamma T$  in top  line of Eq.~\eqref{eq:phi}, we
 obtain
 %
\begin{align}
\Phi(x) =& \frac{\epsilon}{\pi} \int_{1}^{\omega_0/T} \frac{dy}{x+y} \log (y)\,\Phi(y).
\label{eq:int}
\end{align}
The $1/(x+y)$ in the right hand side of this equation is the static part of the spin-mediated interaction in rescaled variables, and  the $\log(y)$ term is the Cooper logarithm from lukewarm fermions, which, we remind, display a Fermi liquid behavior.

     \section{Differential gap equation}
      \label{sec:3}

A convenient way to analyze this equation, applied before to dynamical pairing out of a non-Fermi liquid and to pairing near a van Hove point in a metal~\cite{wang-raghu-torroba-17,paper_1,Ojajarvi2024}
 is to convert the integral equation into a differential one, which is easier to analyze.  For our case, the conversion is done by approximating $1/(x+y)$ in Eq.~\eqref{eq:int} by $1/x$ when $x >y$ and by $1/y$ when $y>x$. This is known as the local approximation~\cite{wang-raghu-torroba-17,paper_1}.
 It is quantitatively exact if  the interaction $V(l)$ is a slow function of the variable $l$.
 In our case the interaction $V(l) \propto 1/l$ is not particularly slow, yet 
 we shall see that the differential gap equation (DGE) remains qualitatively valid.

 Applying this, we obtain from \eqref{eq:int}
   \beq
   \Phi (x) = \frac{\epsilon}{\pi} \left[\frac{1}{x} \int_1^x dy \log(y)\, \Phi (y) + \int_x^{\frac{\omega_0}T} \frac{dy \log(y)}{y}\, \Phi (y)\right].
   \label{ch_4}
   \eeq

   For $x$ well above the lower limit, the second term scales as $\log^2 x$ and the first as $\log {x}$.
   It is then
  tempting to only keep the    second term and and drop the
  first one.
   Imposing this, we obtain  from (\ref{ch_4})
   \beq
   \Phi (x) = \frac{\epsilon}{\pi}  \int_x^{\omega_0/T} \frac{dy \log(y)}{y} \Phi (y).
   \label{ch_5}
   \eeq
  Differentiation over $x$ then yields a local differential equation
  \beq
  \Phi' (x) = -\frac{\epsilon}{\pi}   \frac{\log(x)}{x} \Phi (x)
  \label{ch_6}
   \eeq
   The boundary condition is set by (\ref{ch_5})
    \beq
   \Phi \left(\frac{\omega_0}{T}\right) = 0
   \label{ch_7}
   \eeq
Solving (\ref{ch_6}), we obtain
\beq
\Phi (x) = \Phi_0 \exp\[-\frac{2\epsilon}{\pi} \log^2{x}\].
  \label{ch_8}
   \eeq
Indeed, this is the result of summing the perturbative series in terms of the $\log^2x$ term~\cite{WC}.
However, this function does not satisfy the boundary condition (\ref{ch_7}) at any $T$.
  This implies that the summation of the leading logarithms does not lead to an instability.
  This already shows that the pairing of lukewarm fermions is different from BCS.
  In Appendix \ref{app:1} we further contrast this with color superconductivity, where the gap equation has a similar form, yet
  the summation of the leading logarithms does give rise to a finite $T_c$.

  We now return to the full  Eq.~(\ref{ch_4}) and keep both terms in the right hand side.  One can easily verify that to obtain a local differential equation one has to differentiate twice. Indeed, differentiating one we obtain
   \beq
  \Phi' (x)  =- \frac{\epsilon}{\pi x^2} \int_1^x dy \log(y)\, \Phi (y) =0
  \label{ch_9}
   \eeq
   Differentiating again we obtain
  \beq
  \Phi'' (x)  + \frac{2}{x} \Phi' (x) + \frac{\epsilon}{\pi} \frac{\log(x)}{x^2} \Phi (x) =0
  \label{ch_10}
   \eeq
 The two boundary conditions can be extracted from  the original equation (\ref{ch_4}) and from Eq. (\ref{ch_9}). They are
 \bea
 &&\Phi' (1) =0 ,\nonumber \\
 &&\Phi' \left(\frac{\omega_0}{T}\right) +
 \frac{{T}}{\omega_0}  \Phi \left(\frac{\omega_0}{T}\right) =0.
  \label{ch_11}
  \eea

\subsection{Approximate differential equation}
      \label{sec:3a}

As a first step in the analysis, let's consider an approximate (toy-model) version of Eq.~(\ref{ch_10}) in which we assume that relevant
$x$ are of order the upper limit $x \sim \omega_0/T$, at least to logarithmic accuracy, and replace $\log(x)$ in the right hand side of (\ref{ch_10}) by $\log(\omega_0/T)$.  The differential equation then becomes
\beq
  \Phi'' (x)  + \frac{2}{x} \Phi' (x) +  \frac{D}{x^2} \Phi (x) =0
  \label{ch_12}
   \eeq
where
\beq
D = \frac{\epsilon}{\pi} \log\left(\frac{\omega_0}{T}\right)
 \label{ch_14}
 \eeq
 is the $T$-dependent parameter. At large $T \leq \omega_0$, $D$ is small, of order $\epsilon$. At the smallest $T$, $D$ becomes large.

    The solution of (\ref{ch_12}) is a simple power-law function, which we choose as $\Phi (x) \propto x^{\beta-1/2}$.
  Substituting into  (\ref{ch_12}), we obtain the equation on $\beta$:
  \beq
  \beta^2  + D -\frac{1}{4} =0
  \eeq
   whose solution is $\beta_{1,2} = \pm \sqrt{{1}/{4}-D}$.
    At $D <1/4$ (larger $T$),  $\beta$ is real. The generic solution of (\ref{ch_12}) is then
    \beq
    \Phi (x) = \frac{\Phi_0}{\sqrt{x}} \left(\frac{1}{x^{\beta}} + C x^{\beta}\right).
    \label{ch_15}
    \eeq
    Substituting into the boundary condition \eqref{ch_11} at $x =1$, we fix $C$ to be $C = - (1/2 -\beta)/(1/2 +\beta)$. However, the second boundary condition is not satisfied for any positive $\beta$.  This implies that there is no non-zero solution of  (\ref{ch_12}) when $D <1/4$.

\begin{figure}
\includegraphics[width=\columnwidth]{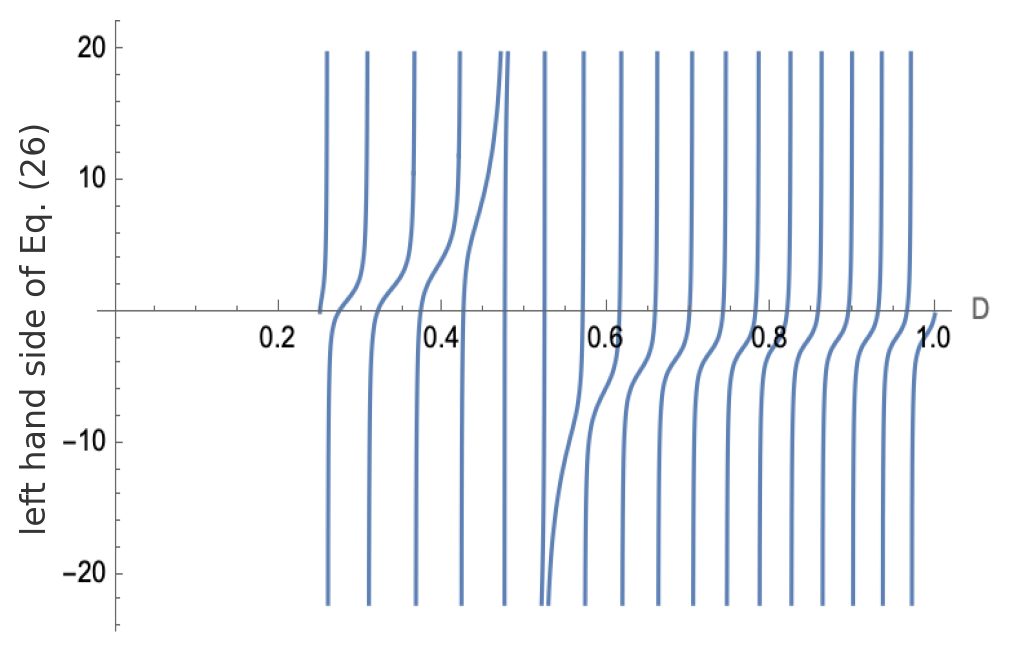}
\caption{Solutions for $D$ in Eq.\eqref{ch_17} for $\epsilon=0.05$, represented by the zeros of the curve. For  $\epsilon\ll 1$, the
 first solution is at $D\approx 1/4.$ }
\label{fig:Q}
\end{figure}

    At $D >1/4$, $\beta$ becomes imaginary $i{\tilde \beta}$, where ${\tilde \beta} = \sqrt{D - 1/4}$.
     A generic solution of (\ref{ch_12}) now becomes
      \beq
    \Phi (x) = \frac{\Phi_0}{\sqrt{x}} \cos\left({\tilde \beta} \log(x) + \phi\right)
    \label{ch_16}
    \eeq
    where $\phi$ is a free parameter.
    Substituting this $\Phi (x)$  into the two boundary conditions in \eqref{ch_11} we find from the condition at $x=1$, $ \tan{\phi} = -1/(2 {\tilde \beta})$, and from the one at $x = \omega_0/T$,
    \beq
    \tan\(\tilde \beta\log{\(\frac{\omega_0}{T}\)} + \phi\) - \frac{1}{2\tilde \beta}=0.
    \label{ex_ch_1}
    \eeq
    Using ${\tilde \beta} = \sqrt{D -1/4}$, Eq. (\ref{ch_14}) for $D$, and the first boundary condition,
    we re-express Eq. (\ref{ex_ch_1})  as
    \begin{align}
    \tan\left(\sqrt{4D -1} \frac{\pi D}{2\epsilon}\right) -\frac{\sqrt{4D-1}}{2D-1} =& 0
     \label{ch_17}
    \end{align}
    The graphical solution of (\ref{ch_17}) is shown in Fig.~\ref{fig:Q}.
    We see that there are infinite number of solutions $T_{c,n}$.  At small $\epsilon$, the solutions are closely packed.
     The largest $T_c$ corresponds to $D\approx 1/4$,\footnote{$D=1/4$ is
      formally also a solution of Eq.~\eqref{ch_17}, but it corresponds to a trivial pairing vertex $\Phi(x)=0.$}
       and is
       given by
    \beq
    T_{c,0} = T_c  \sim \omega_0
    \exp\(-\frac{\pi}{4\epsilon}\)
    \label{ch_17a}
    \eeq
   This $T_c$ is almost the same as in (\ref{ch_2}) except for the factor $\pi/4$ in the exponent.

    We see from this analysis that while there is no threshold  for $T_c$ and $T_c$ exponentially depends on $\epsilon$, like in BCS theory,   other features are  qualitatively different from BCS. Namely, the pairing instability does not emerge from the summation of the leading logarithms, but rather falls into the ``complex exponent" category.     Besides, there is a tower of $T_{c,n}$ for different pairing states.

    \subsection{Full differential equation}
       \label{sec:3b}

We now show that very similar results hold in the original differential equation (\ref{ch_10}) without approximating $\log {x}$ by $\log(\omega_0/T)$.

To this end, it is instructive to
 re-express the derivative over $x$ into the one over $L \equiv \log(x)$. Doing this, we obtain from
 (\ref{ch_10})
\be
\frac{d^2\Phi}{dL^2} + \frac{d\Phi}{dL} + \frac{\epsilon L}{\pi} \Phi = 0.
\label{eq:PhiL}
\ee
The first order derivative above can be removed by defining $\varphi(L) \equiv e^{L/2} \Phi(L)$. The equation for the new function $\varphi(L)$ takes the form of the Airy equation
\be
\varphi''(L) + \left(\frac{\epsilon}{\pi}\,L - \frac{1}{4}\right) \varphi(L) = 0.
\ee
The solution for $\varphi(x)$ can then be straightforwardly expressed as
\begin{align}
\varphi(L) =&C_1\, \mathrm{Ai}(z)+C_2\, \mathrm{Bi}(z),
\label{eq:varphi} \non\\
\textrm{where }
z=& \frac{1}{4}\(\frac{\pi}{\epsilon}\)^{2/3} - \(\frac{\epsilon}{\pi}\)^{1/3}L,
\end{align}
and Ai and Bi are Airy functions of first and second kind.
 Then
 \be
\Phi(L) = e^{-L/2} \[C_1\, \mathrm{Ai}(z)+C_2\, \mathrm{Bi}(z)\].
\label{ex_ch_2}
\ee

\begin{figure}
\includegraphics[width=\columnwidth]{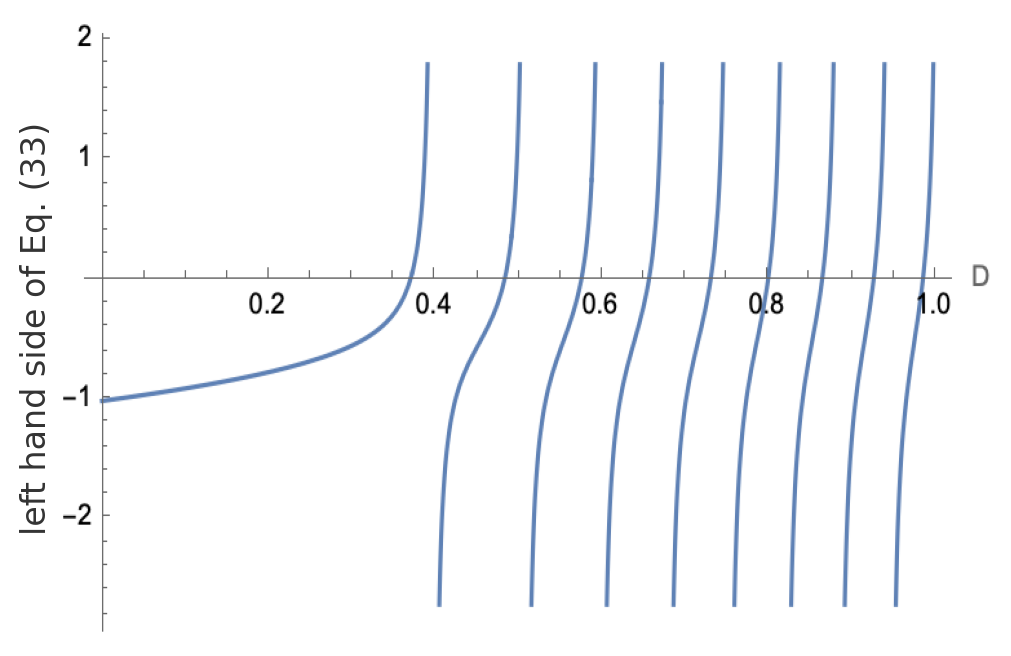}
\caption{Solutions of Eq.~\eqref{ex_ch_4} for $\epsilon=0.05$ in terms of $D$. For small $\epsilon\ll1$, the leading $T_c$ corresponds to $D\approx 1/4.$}
\label{fig:2}
\end{figure}

 The boundary conditions for $\Phi (L)$ are
 \beq
 \Phi' (0) =0,  ~~\Phi'\(\log{\frac{\omega_0}{T}}\) = - \Phi\(\log{\frac{\omega_0}{T}}\)
\label{ex_ch_3}
\eeq
where the derivatives are over $L$.
In the limit $\epsilon\ll 1$, using the asymptotic behavior~\cite{wiki:Airy_function} of the Airy functions that $\operatorname{Ai}'(z)/\operatorname{Ai}(z) \sim -\sqrt{z}$ for $z\gg 1$,  we find that the boundary condition $\Phi'(0)=0$ is satisfied by choosing  $C_2=0$.
The other boundary condition in Eq.~\eqref{ex_ch_3}
yields
\beq
\left(\frac{\epsilon}{\pi}\right)^{1/3}\,\frac{\operatorname{Ai}'(z_D)}{\operatorname{Ai}(z_D)}-\frac{1}{2}=0,
\label{ex_ch_4}
\eeq
where
\beq
z_D= \frac{1}{4}\(\frac{\pi}{\epsilon}\)^{2/3} - \(\frac{\epsilon}{\pi}\)^{1/3} \log\frac{\omega_0}{T}
 = - \left(\frac{\pi}{\epsilon}\right)^{2/3} \left(D -\frac{1}{4}\right)
\eeq
We find the solution of (\ref{ex_ch_4})
graphically in Fig.~\ref{fig:2}. As can be seen, this boundary condition is satisfied at various values of $D>1/4$, corresponding to a tower of $T_{c,n}$'s.  Indeed, from the property of the Airy function~\cite{wiki:Airy_function},  the left hand side of \eqref{ex_ch_4} is  oscillatory for $D>1/4$ and monotonically negative for $D<1/4$. For $\epsilon\ll 1$, the largest $T_{c,0}$ corresponds to $D\approx 1/4$, and thus
\be
T_{c,0} = T_c  \sim \omega_0 \exp\(-\frac{\pi}{4\epsilon}\).
\label{eq:tc}
\ee
We see that the functional form is the same as in (\ref{ch_17a}). However, the sub-leading terms
 in the expansion in $\epsilon$ are different.

In the original variable $x$, the exact solution for $\Phi (x)$ is
\begin{align}
\Phi(x) =\frac{\Phi_0}{\sqrt{x}}\,\mathrm{Ai}\[\frac{1}{4}\(\frac{\pi}{\epsilon}\)^{2/3} - \(\frac{\epsilon}{\pi}\)^{1/3}\log x\],
\label{eq:newPhi}
\end{align}
 We remind the reader that this solution exists only when $D >1/4$.
 Observe that
 Eq.~\eqref{eq:newPhi}
 is similar
 to Eq.~\eqref{ch_16}.

For subsequent analysis,  it is convenient to
 introduce the new function
  \be
~Q(L) = -\frac{d\log \Phi(x)}{d \log x},
\ee
The DGE for $Q(L)$ has the form
       \begin{align}
     Q'(L) =   Q^2(L) - Q(L) +\frac{\epsilon}{\pi} L  ,
        \label{eq:Q}
       \end{align}
       and the boundary conditions are
       \be
      Q(0) = 0,~Q\(\log \frac{\omega_0}{T}\)=1.
      \label{eq:bd}
       \ee
   Eq.~\eqref{eq:Q} is known as a Riccati equation.
    Using (\ref{ex_ch_2}) we obtain its solution:
   %
\be
Q(L)=\frac{1}{2} + \left(\frac{\epsilon}{\pi}\right)^{1/3}\,\frac{\operatorname{Ai}'(z)}{\operatorname{Ai}(z)}.
\label{eq:Qsol}
\ee
The boundary conditions Eq.~\eqref{eq:bd} yield the same tower of $T_c$ as Eq.~(\ref{ex_ch_4}).

\subsection{Computing $T_c$ via algebraic equation for $Q$}
    \label{sec:3c}

We now show that
for small $\epsilon$, the formulas for $T_{c,0}$ and $\Phi(x)$,
consistent with (\ref{eq:tc}, \ref{eq:newPhi}), can be
obtained from (\ref{eq:Q})  in a purely algebraic way, without solving the differential equation.
 For this we note that at
$\epsilon\ll 1$, $Q'$ remains small compared to $Q$ and $Q^2$  in nearly the whole range of $L < \pi/(4 \epsilon)$, except very close to the boundary of this range. Within this range, the DGE  (\ref{eq:Q}) can be
approximated by a quadratic algebraic equation
\be
Q^2(L) - Q(L) +\frac{\epsilon}{\pi} L = 0,
\label{eq:Q1}
\ee

\begin{figure}
\includegraphics[width=\columnwidth]{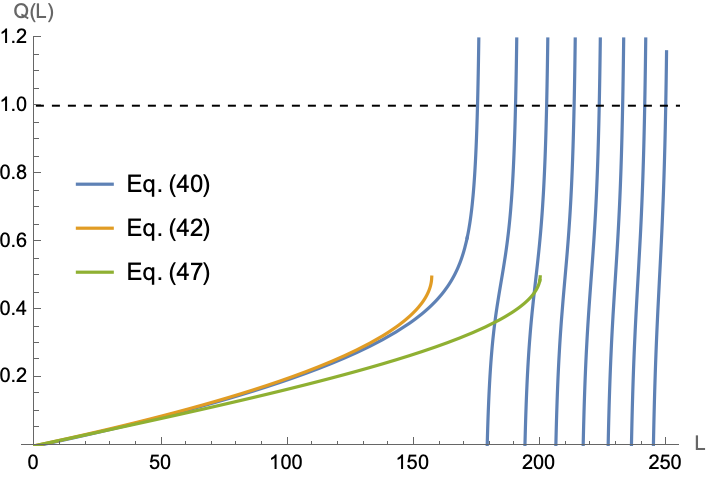}
\caption{Solutions for $Q(L)$ using DGE and algebraic equation with and without improvements beyond local approximation. The boundary condition in Eq.~\eqref{eq:bd} is shown as the dashed line.}
\label{fig:3}
\end{figure}

whose solution is
\be
Q(L)= \frac{1-\sqrt{1-4\epsilon L/\pi}}{2},
\label{eq:alg}
\ee
which we plot in Fig.~\ref{fig:3}.
 We see that the approximation of \eqref{eq:Qsol}
  by (\ref{eq:alg})
  is quite good  for $L<\pi/(4\epsilon)$.
The other quadratic root does not satisfy the boundary condition $Q(0) =0$.
 This solution is valid up to $4 \epsilon L/\pi <1$.
To leading order of $\epsilon$ at a given $L$, this gives
\beq
Q(L) \approx \frac{\epsilon L}{\pi}.
\label{ex_ch_7}
\eeq
The corresponding
\beq
\Phi(x) =\Phi_0 \exp\[-\frac{2\epsilon}{\pi} \log^2{x}\].
\eeq
 is the same as
  in (\ref{ch_8}), i.e., the approximation to linear order in $\epsilon$ for $Q(L)$ is equivalent to solving for $\Phi (x)$ within the leading logarithmic approximation.  The function $Q(L)$ from (\ref{ex_ch_7})
  increases with $L$, but reaches $Q =1/4$ at the maximal $L = \pi/(4\epsilon)$ and then does not satisfy the second boundary condition $Q (L = \log{(\omega_0/T})) =1$.  This is a different way to state that the summation of the leading logarithms in the gap equation does not lead to the pairing susceptibility.

   The full algebraic solution for $Q(L)$ also does not satisfy the boundary condition
   $Q (L = \log(\omega_0/T)) =1$ as its maximum value is $1/2$.  However, there is an important difference with $Q(L)$  from (\ref{ex_ch_7}). Namely, $Q(L)$ from (\ref{eq:alg})
 nearly coincides with  the exact solution of the DGE for all $L < \pi/(4 \epsilon)$ except the ones very close to $\pi/(4 \epsilon)$ (see Fig.~\ref{fig:2}) and, moreover, its derivative diverges at $L \to \pi/(4 \epsilon)$.  Judging from this alone, one can conclude that  the full solution of DGE for $Q' (L)$ should cross
 $Q(L)=1$ at $L$ only slightly above $L = \pi/(4 \epsilon)$.  To find out how this happens one indeed should keep $Q' (L)$ in the DGE, as Fig.~\ref{fig:2} shows.  However, this reasoning already implies that to leading order
  in $\epsilon$, $T_c = T_{c,0}$ can be extracted from the condition $\log{(\omega_0/T_c)}  = \pi/(4 \epsilon)$.
This yields the same $T_c$ as in
\eqref{eq:tc}.  We note, however, that within this reasoning one cannot obtain the tower of $T_{c,n}$.

\subsection{Improved equation for $Q$}
\label{sec:3d}
We now stay within the regime of $L$ where
$Q'\ll Q,Q^2$ and address another issue -- how accurate is the computational procedure leading to
the algebraic equation (\ref{eq:Q1}). We remind that in the derivation of this equation we used local approximation and
 replaced $1/(x+y)$ in the kernel  in Eq. (\ref{eq:int}) by  $1/\max(x,y)$.
We now argue that the local approximation can be improved, at least in the regime where
$Q'$  is smaller than $Q^2$ and $Q$ and can be neglected.
To this end, note that
from the definition of $Q$,
$\Phi (y)$ at a given $y$  can be expressed via $\Phi (x)$ as some $x$ as
\be
\Phi(y)=  \Phi(x) \exp\[-Q(\log x) (\log y - \log x)\],
\ee
where higher-order derivative terms, such as $Q'(\log y)|_{y= x}$,
 have been dropped. Inserting this
 relation into original integral equation Eq.
 (\ref{eq:int}),
we get
\begin{widetext}
\begin{align}
x^{-Q(\log x)}=&\frac{\epsilon}{\pi}\int_1^x \frac{y^{-Q(\log x)}}{x+y}\log y\, dy +\frac{\epsilon}{\pi}\int_x^{\frac{2}{T}} \frac{y^{-Q(\log x)}}{y+x}\log y\, dy\nonumber\\
=&\frac{\epsilon}{\pi}\sum_{n=1}^\infty\frac{(-1)^{n-1}}{x^n}\int_1^x y^{n-1-Q(\log x)}\log y\, dy + \frac{\epsilon}{\pi}\sum_{n=1}^\infty(-1)^{n-1}x^{n-1}\int_x^{\frac{2}{T}} y^{-n-Q(\log x)}\log y\, dy,
\end{align}
\end{widetext}
where in the second line we expanded the kernel in $y$ and $x$ respectively. Only keeping the $n=1$ contribution in both terms amounts to taking the local approximation, which, after performing the integrals,  reproduces Eq.~\eqref{eq:Q1}.  We go beyond local approximation by retaining all terms in the sums.
Evaluating the integrals and performing the re-summation, we obtain a
 more accurate equation for $Q (L)$ in the form
\begin{align}
1=&\frac{\epsilon}{\pi}\left[\sum_{n=-\infty}^{\infty}\frac{(-)^{n+1}}{n-Q(L)}\right]L=\frac{\epsilon L}{\sin\left[\pi Q(L)\right]}.
\label{eq:alg2}
\end{align}
We plot the solution of this equation in Fig.~\ref{fig:3}.
We see that the solution
 gets
  modified from
  Eq.~\eqref{eq:alg} at larger $L$'s, but still only exists in a limited range of $L$, and does not reach $Q(L)=1$ at the boundary of this range. Yet, the solution of (\ref{eq:alg2}) rapidly increases at the edge of its applicability range, much like
 the solution of ~(\ref{eq:alg}).  By analogy, it is reasonable to expect that the largest $L$, where
  the solution of (\ref{eq:alg2}) holds, is close to $\log{\omega_0/T_c}$.
 We see that, in distinction to the solution of (\ref{eq:alg}), the largest $L$ up to which the solution of (\ref{eq:alg2}) holds is $L = 1/\epsilon$.  We then 
 obtain an improved estimate for $T_c$ as
\be
T_c  \sim \omega_0 \exp\(-\frac{1}{\epsilon}\).
\label{ex_ch_8}
\ee
This is the result that we cited in the Introduction.  The formula for $T_c$ looks like the BCS formula, yet, as we demonstrated, the pairing mechanism is not BCS -- it falls into the ``complex exponent" rather than the ``Cooper logarithm" category, and there is a tower of $T_{c,n}$ with topologically distinct $\Phi_n (x)$, similar to pairing out of a non-Fermi liquid. The only distinction with the latter is the absence of the threshold
for the coupling, which results from the coherent Fermi-liquid behavior of lukewarm fermions.

 Eq.~(\ref{ex_ch_8}) was  first obtained 
 by us in Ref.~\cite{WC} and was found to
 match numerics quite well. In that paper, we, however, did not explore the fact that one needs to keep $Q'$ in the DGE in order to satisfy the boundary condition $Q(\log(\omega_0/T)) =1$ and did not emphasize the non-BCS nature of the pairing of lukewarm fermions.
It is tempting to further improve the DGE by
incorporating $Q'$ terms into Eq.~\eqref{eq:alg2}. This, however, requires rather sophisticated calculations outside of local approximation.
We leave this issue for future work.

\section{RG formulation of the pairing problem and its relation to DGE}
    \label{sec:4}

In this section, we reformulate the pairing problem using the technique of momentum-shell RG, and show that it is fully equivalent with the differential equation.

Within the BCS pairing mechanism, under the RG flow toward the FS, the tangential component of momentum $k_\|$ does not rescale~\cite{shankar,polchinski}. However, in our problem, since the self-energy for lukewarm fermions explicitly depend on $k_\|$, it is crucial that $k_\|$ also rescales under RG. Given the form of the self-energy, we employ an RG scheme with the following scaling dimensions:
\be
[k_\|] = \frac{1}2,~[k_{\perp}] = \frac{1}{2},~[\omega] = 1.
\ee

As the instability arises from integrating over $\kpara$, we implement a momentum-shell RG process during which at each step, (i) modes in a momentum shell $\kpara^2\in(\Lambda (1-D \ell),\Lambda)$ are integrated out, and then (ii) the upper cutoff is formally rescaled back to $\Lambda$.

As an immediate consequence of $[k_{\|}]\neq 0$, the engineering dimension of the pairing coupling $g$ is altered. Unlike the BCS pairing mechanism in which $g$ is marginal~\cite{shankar,polchinski}, here $g$ is irrelevant, and by a similar analysis to that in Ref.~\cite{polchinski}, we get
\be
[g] = -1.
\ee
At each RG step, the scaling dimension of $g$ leads to
\be
\delta_1 g = - g \delta \ell.
\label{eq:8}
\ee

There are two additional fundamental differences with the BCS pairing mechanism. First, the coupling ``constant'' $g$ is not a constant, but rather by itself a function of $k_{\|}$, i.e., explicitly scale dependent. As a common feature in all quantum-critical pairing models, the standard treatment within RG~\cite{color1,raghu2015metallic} is to identify $g$ with the pairing interaction at a given running scale $\Lambda$. This is equivalent with the local approximation, and is technically speaking accurate (to log accuracy) if the pairing interaction is slow-varying at log scale. This is satisfied by the $\gamma$-model at small $\gamma$ and color superconductivity in $3+1$d. In our case, the pairing interaction $D(k_{\|})\sim \epsilon/(\pi k_{\|}^2)$, and such treatment is not quantitatively accurate. Still, as we shall see, it correctly captures the qualitative feature of the pairing mechanism. The local approximation leads to an inherent running of $g$ aside from regular contributions from the RG procedure. This leads to, during an RG step $\Lambda\to \Lambda(1-\delta\ell)$:
\be
\delta_2 g =  \frac{\epsilon}{\pi} \frac{\delta \ell}{\Lambda}.
\label{eq:9}
\ee

Second, the 1-loop contribution to the running of $g$ is also different from BCS. In BCS, the 1-loop contribution to the RG flow comes from a shell of $(k_{\perp},\omega)$ with $\kpara$ integrated over. Here, $\kpara$ is the running scale. In a momentum shell $\kpara^2\in(\Lambda (1-\delta \ell),\Lambda)$, the contribution to 1-loop pairing susceptibility is
\be
g^2 \int_{\Lambda(1-\delta \ell)}^{\Lambda} d\kpara^2\int_{Te^\ell}^{\kpara^2} \frac{d\omega}{\omega}.
\ee
Note that the lower cutoff of the frequency integral is rescaled to $Te^\ell$, instead of $T$, due to the sub-steps (ii) during the RG flow ($\ell = \int \delta \ell$). This leads to a change in $g$ given by
\be
\delta_3g= g^2\Lambda\(\log\frac{\Lambda}{T} - \ell\)\delta \ell.
\label{eq:11}
\ee

Combining Eqs.~(\ref{eq:8}, \ref{eq:9}, \ref{eq:11}), we get the following $\beta$-function for the RG flow of $g$:
\be
\frac{dg}{d\ell} = \frac{\epsilon}{\pi \Lambda} -g + g^2 \Lambda \(\log\frac{\Lambda}{T} - \ell\).
\label{eq:g}
\ee
The RG flow begins at the UV scale $\Lambda = \omega_0$ and stops at  $\ell=\log(\omega_0/T)$. For the pairing instability at $T=T_c$, we have the boundary conditions
\be
g(\ell=0) = \frac{\epsilon}{\pi\Lambda},~~~g\(\ell=\log\frac{\omega_0}{T}\) \to \infty.
\label{eq:gb}
\ee

Remarkably, if one identifies $L$ in \eqref{eq:Q} with the ``reverse RG time'', and $Q$ with the ``inverse coupling''~\cite{wang-raghu-torroba-17}, i.e.,
\be
L \leftrightarrow \log\frac{\Lambda}{T} -\ell,~~~Q(L)\leftrightarrow \frac{\epsilon \Lambda}{\pi g(\ell)},
\ee
the RG equation \eqref{eq:g} becomes exactly the same as the DGE,
Moreover, the boundary conditions for $g$ in Eq.~\eqref{eq:gb} precisely maps to those for $Q$ in Eq.~\eqref{eq:bd}.

For a number of pairing problems, it is known that the approaches of Eliashberg equation and RG are equivalent~\cite{color1,wang-raghu-torroba-17}. In particular, for pairing mediated by dynamic interactions, Ref.~\cite{wang-raghu-torroba-17} showed explicitly that the differential equation converted from the Eliashberg integral equation can be directly mapped to an RG equation. For the present problem, the  pairing interaction is predominantly static, but the relation between $\Phi(\kpara)$ and $g(\ell)$ still holds. For completeness, in Appendix \ref{app:1}  we explicitly demonstrate that for color superconductivity mediated by massless gluons and for the $\gamma$-model, the RG equations and the differential equations from Eliashberg equations are exactly the same.

The derivation of the RG equation makes clear the fundamental difference between our pairing mechanism and BCS. For the BCS mechanism, the RG equation is simply
\be
\frac{dg}{d\ell} = g^2,
\ee
in which the fermionic self-energy and the form of the pairing interaction does not play an important role.

The last term in Eq.~\eqref{eq:g} is quite similar to the RG equation for pairing in graphene at van Hove doping~\cite{graphene_vH}, whose one-loop pairing susceptibility contains $\log^2$ just like our problem.
Had we only kept this term, solving the RG equation would yield $T_c = \omega_0 \exp (-\sqrt{2/{\epsilon}})$, quite like that for van-Hove doped graphene~\cite{graphene_vH} and for color superconductivity~\cite{color1,CS} (see Appendix \ref{app:1a}). However, the additional terms in the RG equation lead to a different, BCS-like formula for $T_c$, in spite of a fundamentally different pairing mechanism.

Compared with the RG equation for the $\gamma$-model~\cite{raghu2015metallic} (see Appendix \ref{app:1b}), the first two terms in Eq.~\eqref{eq:g} are similar. The only difference is the additional $L$ factor in the last term. As a direct consequence, there is no  fixed point at $\ell\to\infty$ for an arbitrarily small $\epsilon$, i.e., our pairing problem does not have a threshold for pairing strength.

    \section{Conclusions}
   \label{sec:5}
  In this work, we argued that the very appearance of BCS-looking formula for superconducting $T_c$
   does not necessary imply BCS pairing mechanism, rooted in the geometric series of the Cooper logarithms.
   We considered as an example pairing of lukewarm fermions in the critical spin-fermion model.
     These fermions are located near hot spots, but still at some distance from them, and their self-energy
      has a Fermi liquid form, but $d\Sigma (\omega,k)/d\omega \propto 1/|k_\parallel|$, where
      $k_\parallel$ is the distance of a lukewarm fermion to a hot spot.  We showed using several computational procedures, with varying degrees of
       accuracy,
       that lukewarm fermions pair already  for arbitrary weak attraction $\epsilon$, and the corresponding $T_c$ is exponential in $1/\epsilon$, like in BCS theory.
      At a first glance, this looks like BCS pairing.  However, we demonstrated that the pairing mechanism is qualitatively different  from BCS in three aspects: (i) the summation of the leading logarithms does not lead to pairing, (ii) pairing comes subleading
      terms and develops when the exponent for the pairing susceptibility becomes complex, and (iii) there is an infinite set of critcal temperatures for topologically distinct gap functions.   All three properties are characteristics of fundamentally non-BCS pairing. From a field-theoretic perspective,
      the distinction between the present pairing mechanism and BCS is rooted in their qualitatively different  behaviors of pairing coupling constant under RG flow.

\begin{acknowledgments}
We are grateful to D. Maslov, J. Schmalian, G. Torroba, and H. Wang for helpful discussions. This work was supported by
National Science Foundation grant NSF DMR-2045781 (YW) and NSF DMR-2325357 (AVC).
 The work was initiated  when
YW and AVC visited
KITP at UCSB.  KITP is supported in part
by the National Science Foundation under PHY-1748958.
\end{acknowledgments}

\appendix
\section{Equivalence between DGE and RG equation for the running coupling}
\label{app:1}

In this Appendix we analyze the connection between the differential equation for the pairing vertex and the RG equation for the running coupling.

A differential equation for the pairing vertex is obtained from the original integral Eliashberg equation for the pairing vertex by taking the local approximation, which  simplifies the pairing interaction $V(a-b)$  where $a$ and $b$  are external and internal  variables (momentum or frequency), to $V(a)$ for $a >b$ and $V(b)$  for $b >a$,  and the RG equation for the coupling is obtained by using Wilsonian coarse-graining procedure with the convention that the RG coupling has to diverge  at the energy scale set by $T_c$.

 A generic analysis of the interplay has been performed in Ref.~\cite{wang-raghu-torroba-17}, where the authors demonstrated that for a generic dynamical interaction $V(a-b)$,  $T_c$ extracted from the differential equation and from RG are equivalent.

We focus on two specific examples --- color superconductivity (pairing by a logarithmically singular dynamical interaction) and the pairing by the effective dynamical interaction $V(\Omega_m) \propto 1/|\Omega_m|^\gamma$ (the $\gamma$ model).  For both cases we present the explicit expressions for the
 differential equation and its solution, and convert the equation for the pairing vertex $\Phi$ into the one for the RG coupling $g$.

\subsection{Color superconductivity}
\label{app:1a}
Color superconductivity occurs in quark matter due to condensation of quark pairs (diquarks) driven by logarithmically singular attractive interactions mediated by gluon exchange~\cite{color1}. Despite its exotic nature, the same pairing mechanism  can also be realized in electronic pairing, as long as pairing interaction is logarithmically singular (e.g., interaction mediated by nematic fluctuations in 3d.)
 The instability temperature for color superconductivity has been  found by D. Son within the RG framework.~\cite{color1}  (see also \cite{son_1,*son_b,*son_3}).
 Chubukov and Schmalian~\cite{CS,*CS_1}
  reproduced Son's result for $T_c$ by constructing Eliashberg equation for the pairing vertex with a source term (an integral equation in frequency) and analyzing the pairing susceptibility.

  The point of departure for our analysis is the integral Eliashberg equation in Matsubara frequencies,  without the source term. For logarithmic interaction it has the form
  \beq
  \Phi (x) = \lambda \int_{\bar T}^1 \frac{d y}{y} \log{\(\frac{1}{|x-y|}\)} \Phi (y)
  \label{a_1}
  \eeq
   where $x, y >0$ are re-scaled frequencies $x = \omega_m/\Lambda$, where $\Lambda$ is the upper cutoff for logarithmical behavior of the interaction, ${\bar T} = T/\Lambda$, and $\lambda$ is a dimensional coupling, which we assume to be small.

   The differential equation corresponding to (\ref{a_1}) is  obtained using the same procedure as in the main text. It is
  \beq
  \Phi'' (x) + \frac{\Phi' (x)}{x} + \lambda \frac{\Phi (x)}{x^2}=0,
  \label{a_2}
  \eeq
  and the boundary conditions are
  \beq
  \Phi (x=1) =0, ~~ \Phi' (x = {\bar T}) =0
  \label{a_3}
  \eeq
  The solution of (\ref{a_2}) is a power-law, $\Phi (x) \propto x^\beta$.  Substituting into (\ref{a_2}), one finds that
  $\beta = \pm i \sqrt{\lambda}$ is imaginary for any non-zero $\lambda$.  Accordingly,
  \beq
  \Phi = \Phi_0 \cos{\left(\sqrt{\lambda} \log{\(\frac{1}{x}\)} + \phi\right)}
  \label{a_4}
  \eeq
where $\phi$ is a free parameter.  Substituting into (\ref{a_3}) we find $\phi = \pi/2$ and
\beq
\sqrt{\lambda} \log{\(\frac{1}{\bar T}\)} = \frac{\pi}{2} (1 + 2n)
 \label{a_5}
  \eeq
Hence
\beq
T_{c,n} \sim \Lambda \exp\[-\frac{\pi (1+2n)}{2 \sqrt{\lambda}}\]
\label{a_6}
  \eeq
 Analyzing the solution, we  find that the case of color superconductivity is an intermediate between BCS and non-BCS pairing and has features of both.  Like in BCS case, the pairing does come from the summation of the leading logarithms (series of $\log^2$ terms), and $T_c$ is exponential in $1/\sqrt{\lambda}$. Like in non-BCS case, the pairing instability is associated with the complex exponent in the power-law solution for $\Phi (x)$, and there is a tower of $T_{c,n}$ for topologically distinct $\Phi_n (x)$ with $n$ zeros on a positive Matsubara semi-axis.

We now demonstrate how to convert Eqs. (\ref{a_2}), (\ref{a_3}) into the equation for the running  coupling $g$ that departs from a constant at $x=1$ and diverges at $ x = {\bar T}$.
 For this we follow the analysis in Sec.~\ref{sec:4}
  of the main text and introduce
  $p = \log{1/x}$ and $Q = -d (\log{\Phi (x)})/{d \log(x)} = d(\log{\Phi (p)})/dp$.
  Re-expressing  (\ref{a_2}) as the equation on $Q(p)$, we find
 \beq
 Q' (p) + Q^2 (p) + \lambda =0
 \eeq
  where the derivative is with respect to $p$, and $p$ is running between $p_\textrm{min} =0$ and $p_\textrm{max} = \log(\omega_0/T)$.
  The boundary conditions are
 \beq
 Q(p_\textrm{min})= \infty,~~Q(p_\textrm{max}) = 0
 \eeq
 Introducing then $g(p) =1/Q(p)$, we finally obtain
  \beq
 g' (p)  = 1 + \lambda g^2
\label{a_7}
 \eeq
with the boundary conditions
\beq
 g(p_\textrm{min})= 0,~~g(p_\textrm{max}) = \infty
 \label{a_8}
 \eeq
 This coincides with the equation for RG coupling, obtained by Son~\cite{color1}. Solving (\ref{a_7}) and using (\ref{a_8}) one obtains
 \beq
 g(p) = \frac{1}{\sqrt{\lambda}} \tan{\left(\sqrt{\lambda p}\right)}
 \eeq
   subject to $\cos{\left(\sqrt{\lambda p_\textrm{max}}\right)} =0$.  This leads to the same tower of $T_{c,n}$ as in (\ref{a_6}).

\subsection{$\gamma$-model}
\label{app:1b}
 We now apply the same reasoning to the pairing in the $\gamma$-model~\cite{acf1,first-mats-1,Yuzbashyan1}, which describes a set of quantum-critical models with the effective interaction $V(\Omega_m) \propto 1/|\Omega_m|^\gamma$ and  normal state  self-energy $\Sigma (\omega_m) = |\omega_m|^{1-\gamma} \omega^\gamma_0 \sgn (\omega_m)$, where $\omega_0$ is a function of  the coupling between fermions and a critical boson.

 As before, we  use the integral gap equation in frequency as the point of departure of our analysis.
  It has been presented and discussed in Ref.~\cite{paper_1,paper_2}, where the authors also analyzed the full corresponding differential equation.
  For our purpose of comparing with RG, it is sufficient to set  the lower limit of  frequency integration  at $T$ and the upper one at $\omega_0$ and neglect bare $\omega$ compared to the self-energy.  We also restrict with $\gamma <1$.  The truncated Eliashberg equation is
  \beq
  \Phi (x) = \lambda \int^1_{\bar T} \frac{dy}{|x-y|^\gamma} \frac{\Phi (y)}{y^{1-\gamma}}
  \label{a_9}
  \eeq
 where, as before, $x, y >0$ are frequencies in units of $\omega_0$, ${\bar T} = T/\omega_0$, and $\lambda$ is a dimensional coupling related to the ratio of fully dressed interactions in the particle-particle and particle-hole channels (in the extended matrix $\mathrm{SU}(N)$ model with $N \gg1$, $\lambda \propto 1/N$~\cite{raghu-enhanced,wang-wang-torroba}).
 Converting Eq.~(\ref{a_9}) into the differential equation in the same way as in the main text and introducing $z = x^\gamma$, we obtain
  \beq
  \Phi'' (z) + \frac{2}{z} \Phi' (z) + \frac{\lambda}{\gamma} \frac{\Phi (z)}{z^2} = 0
  \label{a_10}
  \eeq
  with the boundary conditions being
  \beq
  \Phi' (1) + \Phi (1) =0, ~~ \Phi' (z = {\bar T}^\gamma) =0
  \label{a_11}
  \eeq
  The derivatives are with respect to $z$.
 The solution is again a power-law $\Phi (z) \propto z^{\beta-1/2}$ ($-1/2$ is added to simplify formulas below).
  Substituting into (\ref{a_10}), one obtains
  \beq
  \beta = \pm \frac{1}{2} \sqrt{1-\frac{4\lambda}{\gamma}}
  \label{a_12}
  \eeq
  For small $\lambda$, $\beta$ are real.  A simple experimentation shows that one cannot satisfy the two boundary conditions in (\ref{a_11}). (For more detailed discussion on this issue, see Refs.~\cite{paper_1,zhang2025}.)
 This implies that there is no pairing instability at weak coupling.

 \begin{figure}[t]
    \includegraphics[width=\columnwidth]{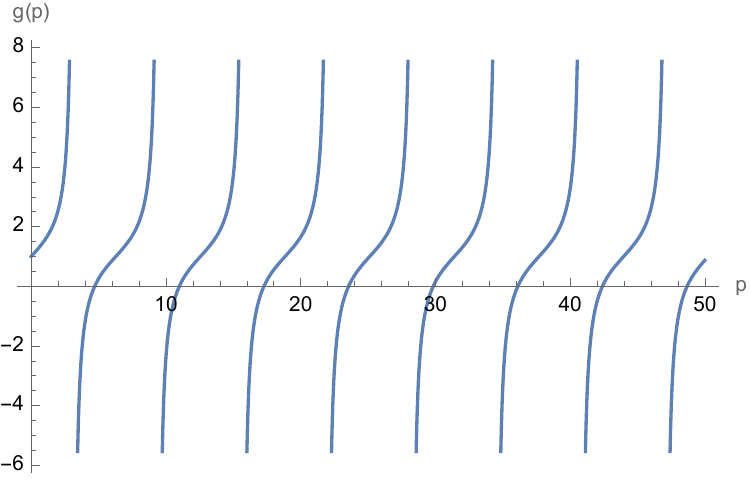}
    \caption{The plot of $g(p)$ for $a=2$.    There is an infinite set of $p \sim \log{(\omega_0/T)^\gamma}$, at which $g(p)$ diverges. }
    \label{fig:gp}
\end{figure}

 At larger $\lambda > \lambda_c = \gamma/4$, $\beta = i {\tilde \beta}$ is imaginary (${\tilde \beta} = 0.5 \sqrt{\frac{\lambda}{\lambda_c}-1}$).   Now
 \beq
 \Phi (z) = \frac{1}{\sqrt{z}} \cos{\left({\tilde \beta} \log(z) + \phi\right)}
  \label{a_14}
  \eeq
 where $\phi$ is arbitrarily.
 Substituting into (\ref{a_11}) we obtain
 \beq
 \tan{\phi} = -\frac{1}{2{\tilde \beta}},~~~ \tan{\left({\tilde \beta} \log{\frac{1}{{\bar T}^\gamma}}\right)} = \frac{2 {\tilde \beta}}{2 {\tilde \beta}-1}
  \label{a_141}
  \eeq
 Solving (\ref{a_141})
  for $\lambda \geq \lambda_c$,
   we obtain a tower of critical temperatures~\cite{paper_1}
\beq
T_{c,n} \sim \omega_0 \exp\(\frac{2\pi (1+n)}{\gamma} \sqrt{\frac{\lambda_c}{\lambda-\lambda_c}}\)
 \label{a_15}
 \eeq
 where $n=0, 1, 2...$. Like before, each $T_{c,n}$ is the onset temperature for a topologically distinct $\Phi_n (z)$ with $n$ zeros on a positive Matsubara semi-axis.

 To convert (\ref{a_10}) into the equation for the running coupling $g$ that departs from a constant at $z=1$ and diverges at $ z = {\bar T}^\gamma$, we again follow the analysis in Sec. \ref{sec:4}
  of the main text and introduce
  $p = \log{1/z}$ and $Q = - d (\log{\Phi (z)})/{d \log(z)} = d(\log{\Phi (p)})/dp$.
  Re-expressing  (\ref{a_10}) as the equation on $Q(p)$, we find
 \beq
 Q' (p) + Q^2(p) - Q (p) + \frac{\lambda}{\gamma} =0
 \eeq
  where the derivative is with respect to $p$, and $p$ is running between $p_\textrm{min} =0$ and $p_\textrm{max} = \log{(\omega_0/T)^\gamma}$.
  The boundary conditions are
 \beq
 Q(p_\textrm{min})= 1,~~Q(p_\textrm{max}) = 0
 \eeq
 Introducing as before $g(p) =1/Q(p)$, we finally obtain
  \beq
 g' (p)  = 1 -g + \frac{\lambda}{\gamma} g^2
\label{a_16}
 \eeq
with the boundary conditions
\beq
 g(p_\textrm{min})= 0,~~g(p_\textrm{max}) = \infty.
 \label{a_17}
 \eeq

 One can verify that (\ref{a_8})  coincides with the equation for RG coupling, derived in Ref.~\cite{raghu2015metallic}.
 Solving (\ref{a_16}) and using (\ref{a_17}) one obtains
 \begin{align}
 g(p) = \frac{2}{a} \Bigl\{&1 + \sqrt{a-1} \Bigr.\label{a_18}
\\
 &\Bigl.\tan{\left[\frac{1}{2}\left(\sqrt{a-1} p + 2 \arctan{\frac{a-2}{2\sqrt{a-1}}}\right)\right]}\Bigr\}\non
\end{align}
  where $a = \lambda/\lambda_c >1$. This $g(p)$ is subject to $g(p = \omega_0/T^\gamma) = \infty$.
   We plot $g(p)$ in Fig.~\ref{fig:gp}.
    We see that there is an infinite set of points where $g(p) = \infty$, hence a tower of $T_{c,n}$.
    A simple analytical analysis shows that for $\lambda \geq \lambda_c$, these temperatures coincide with $T_{c,n}$ in (\ref{a_15}), as they should.
    
    \section{The relation between present work and Ref. \cite{WC}}
\label{app:2}

As we said in the main text, the two us  analyzed  pairing of lukewarm fermions back in 2013.
We found BCS-looking formula for $T_c$ and indirectly suggested that this implies that the whole pairing problem is BCS-like. Our argument was that lukewarm fermions retain Fermi liquid behavior even at a QCP.

In this communication,  we corrected ourselves and demonstrated that the pairing problem is in fact qualitatively different from BCS. The present argument is argument that  the self-energy of this fermions has a singular dependence on momentum distance from a hot spot.
Because of this singular dependence, the gap equation after frequency integration becomes a integral equation over momentum with a singular kernel.  This, we argued,  makes the pairing problem non-BCS like and similar to other problems of pairing out of a non-Fermi liquid. However, because the singularity of the kernel is only logarithmical, 
  the coupling, which we labeled as $D$ (see Eq.~\eqref{ch_12}), contains the product of $\epsilon$ and  $\log{(\omega_0/T)}$.
  As a signature of non-BCS pairing, the solution of the linearized gap equation exists at a non-zero threshold value $D_c$ of order one (there is a tower of solutions, but let's focus here on the smallest  $D_c$. It is still of order one).
   The condition on $T_c$ is then $\epsilon \log{(\omega_0/T)} = O(1)$. This yields BCS-like formula for $T_c$, the same as we obtained in Ref.~\cite{WC}. However, the pairing mechanism is not BCS. We have also confirmed this by a RG analysis. We re-iterate that this is a consequence of singular momentum dependence of the self-energy for a lukewarm fermion.

\bibliography{ref.bib}
\end{document}